# A renewed search for water maser emission from Mira Variables

**B. M. Lewis**

Arecibo Observatory

Arecibo Observatory,
PO Box 995
Arecibo, PR00613
blewis@naic.edu

## Abstract

There is an ≈60% detection rate for 1612 MHz masers in association with red, color– selected IRAS sources, though few are detected from the bluer circumstellar shells of Mira variables. On the other hand and complementarily, past, pre-IRAS 22 GHz surveys detected many water masers in association with Mira variables. This paper reports on a 22 GHz survey of blue, color-selected Miras at Haystack, wherein 18 new detections are found from 238 searched objects.

## 1. Introduction

The dusty circumstellar shells surrounding late-type stars are copious sources of far infra red radiation. These are readily identified by color-selection from the IRAS *Point Source Catalogue* [PSC]. Translucent shells are usually associated with Mira variables, whereas optically thick shells are often associated with OH/IR stars (Olnon *et al.* 1984). Since good source positions are given in the PSC, it was quickly appreciated that new OH/IR stars could be efficiently identified by detecting 1612 MHz masers at the positions of red, color-selected sources (Engels *et al.* 1984; Lewis *et al.* 1985). This process only needs the deployment of appropriate integration time on-source to detect 1612 MHz emission from ≈60% of them (e.g. Eder *et al.* 1988), whereas earlier discoveries of OH/IR stars required the 1612 MHz emission to be intense enough to be detected when it was scanned by a telescope beam. Systematic searches of color-selected sources have now resulted in the identification of more than 1400 new OH/IR stars (Lewis 1996). But only a few Mira variables are associated with 1612 MHz masers.

Circumstellar shells (CSs) often host SiO, water, and mainline OH masers (Engels 1979; Benson *et al.* 1990) as well as 1612 MHz masers, though present statistics show appreciable diversity about which masers occur in which sources. Nevertheless, water masers are frequently associated with shells of every IR color, including both OH/IR stars and Mira variables. Lewis (1989, 1990, 1996) correlates the detectability of 22 GHz water maser emission with the presence of other masing species, and with the IR colors of shells. OH / IR stars discovered by color–selecting IRAS sources, which are at larger distances on average than most surveyed Miras, have a 55% detection rate for 22 GHz masers (Lewis & Engels 1988, Engels & Lewis 1996). However, these surveys benefit from known positions, known velocities, and the guarantee of only being concerned with oxygen–rich shells. Yet existing data suggest that **all** oxygen-rich Miras have CSs (Lewis 1990), and that most should exhibit water masers if repeatedly observed, so the inherent variability of 22 GHz emission is countered. This is suggested by the 75% detection rate from Miras within 400 pc of the Sun (Bowers & Hagen 1984), where the relatively low sensitivity of the then extant searches is minimized. But the net detection rate from all surveys of stars prior to the advent of IRAS is <26%: it was 17% in the Bowers & Hagen (1984) Mira investigation, 9% in one survey from Haystack (Crocker & Hagen 1983) and 3.4% in one from Green Bank (Dickinson & Dinger 1982).

The IR colors of circumstellar shells range seamlessly from those of the most translucent about well known Mira variables, to those of the most opaque around OH/IR stars. Since 1612 MHz masers are most generally supported by the thicker shells, while 22 GHz masers have a broader color distribution that runs across both the Miras and the OH/IR stars, the characteristics of shells that enable them to support masers is tracked better by 22 GHz than by 1612 MHz observations, particularly at blue IR colors. In 1988, when this survey was conceived, I wondered whether 22 GHz detection results would improve if sample selection was guided by the IRAS parameters of blue CSs with known masers, by analogy with the process that was so effective at finding new OH/IR stars. It was already clear at that time that the IRC sources with the IR colors of photospheres could be readily distinguished and excluded from samples. This paper presents the results of an observing run using the Haystack 120 ft antenna to test the utility of this approach.

## 2. Sample

In January 1988 I assembled a database of 22 GHz water observations, which covered most of the 22 GHz surveys then extant. In particular this included the objects listed in Engels (1979), as well as those in the surveys of Bowers & Hagen (1984), Crocker & Hagen (1983), and Dickinson & Dinger (1982). These surveys were all made with ≈130 ft-class telescopes. The characteristics of detected and nondetected objects were then assessed with respect to their IR parameters, the key data being the colors, IR spectral type, and the magnitude of the 25 μm flux, S(25). The IR colors are defined in the $\nu S_\nu$ formalism as ratios of the 12, 25, & 60 μm PSC fluxes, adjusted for a 300 K blackbody temperature, such that

$$(25 - 12)\ \mu m = \log_{10} \{S(25) * 12 * 0.89 / S(12) * 25 * 1.09\}$$

and $$(60 - 25)\ \mu m = \log_{10} \{S(60) * 25 * 0.82 / S(25) * 60 * 0.89\} \qquad (1).$$

These colors make it easy to distinguish IR luminous photospheres from objects with dusty circumstellar shells. Thus Hacking *et al.* (1985) find a clear separation between them and both carbon and oxygen-rich Miras, while Lewis (1990) finds that almost no objects with photospheric IR colors exhibit circumstellar masers.

The IR color range for the present sample is set on this basis to $-0.9 < (25\text{-}12)\ \mu m < -0.6$, with the blue bound determined by the onset of CSs and the red by the high frequency of 1612 MHz detections. The sample was also limited to objects with S(25) > 7 Jy, as a consequence of a correlation between the frequency of detected 22 GHz emission and S(25). But a substantial fraction of nondetections in the database satisfy these criteria, and many are known to be O-rich Miras. This suggests that the inherent variability of water masers may be a major factor in their detection. The present sample of 238 objects is accordingly dominated by the 221 satisfying the above criteria that had not previously been detected when observed. These are augmented by 17 (from a list of 27) having S(25) > 50 Jy that were known to have SiO and/or mainline OH masers, which had never been searched at 22 GHz.

These Haystack observations were made between 5-10th May 1989, when I searched for 22.235080 GHz emission from the $6_{16}$-$5_{23}$ transition of water. A basic observational unit consists of a single 15 minute, frequency-switched observation covering a ±50 km $s^{-1}$ velocity range about the center, where a correlator lag is 0.22 km $s^{-1}$ wide. Subsequent followup observations either repeated the first to seek confirmation of possible features, or more usually were set at ±80 km $s^{-1}$ about the first to extend the velocity range of coverage. Both the pointing and intensity scale were calibrated against W3(OH): spectra were corrected for atmospheric and elevation-dependent attenuation using the default Observatory procedures that utilise the current temperature and dew point. A 10 Jy $K^{-1}$ conversion factor was applied to the data. The weather was always overcast, usually moist, so the system temperature was generally larger than 100 K.

## 3. Results

This is a detection survey, with a usual $3\sigma$ > 1 Jy, so few objects with peak intensities <1 Jy were detected. Table 1 lists the 18 new detections whose spectra are presented in Fig. 1. Table 1 also includes R Hya, which was first detected by Bowers & Hagen (1984), and EG Lib, which is included as an uncertain detection since it has a rather messy spectrum. But this still led Lewis *et al.* (1995) to find 1665 & 1667 MHz masers at much the same velocity. The nondetections from this survey are listed in Table 2 together with the searched velocity range and their current water-maser status. The net detection rate here is 18/237 or 7.6%. However when account is taken of 42 subsequent detections, most from unpublished Effelsberg observations, the detection rate from the whole sample stands at the moment at 26% (61/238).

The new detections have been searched for mainline OH emission with the exception of BE Phe, which is too far South to be reached from Nançay; many also have unpublished 22 GHz observations from Effelsberg. Table 1 lists references to other detections that validate features seen here. Yet 7 objects have no confirmation of this kind, though Benson *et al.* (1990) do list compatible stellar velocities for UX And and Y Cen of 8.2 and 6 km $s^{-1}$ respectively. The situation for BE Phe is confused as SiO detections by Hall *et al.* (1990) and Haikala (1990) agree with each other in providing a $v_* \approx 18$ km $s^{-1}$, whereas the water and CO detections (Nyman *et al.*1992) both suggest a $v_* \approx -100$ km $s^{-1}$. The remaining objects have no confirmation as yet, though they are all > 5σ detections.

Figure 2 shows the color distribution of the sample, while Figure 3 shows its dependence on S(25). There is not much of a color signature in the detections, but there is a tendency for a higher detection frequency from the objects with the larger S(25) fluxes: of those with S(25) > 20 Jy, 9% (12/134) are detected here, whereas just 7/104 of the rest are. This difference is amplified when calculated from the present state of detections in this sample, as 31% (41/134) of those with S(25) > 20 Jy have been detected compared to 19% (20/104) otherwise.

This program began with a list of 27 objects having (i) known velocities, (ii) S(25) > 50 Jy, that (iii) had not previously been searched for water. Only BE Phe was detected from the 17 observed here, which gives them a detection rate of just 6%. However, later observations have detected 4 more of this subset, to give it a current detection rate of 29%, which is little different from the comparable rate from the whole sample. Interestingly though, of the 10 objects in the subset that were not examined here, 8 have since been detected. The subset members that were observed here are shown with an asterisk in Table 2.

## 4. Discussion

With the exception of BE Phe, the detections here had all been observed unsucessfully before, so they are largely a consequence of the variability of water masers as these briefly rise above the detection threshold. In this context it is worth noting that fifteen of the nondetections were searched independently at Haystack by Benson *et al.* (1996) with similar sensitivity to that deployed here, on average 5.3 times: just 2 were detected. They searched BM Gem 20 times and TX Cam 15 times without success, though TX Cam was detected at a velocity of +9.7 km $s^{-1}$ and an intensity of 0.3 Jy in March 1991 at Effelsberg. Four other Table 2 objects in common with Benson *et al.* (X Gem = 06439+3019, W Cnc = 09069+2527, RS Lib = 15214-2244, & UY Oph = 17139+0446) have Effelsberg intensities <1 Jy. Indeed just 16 of the 42 subsequent detections among our nondetections had peak fluxes >1 Jy, which would enable them to be detected at Haystack, while 11 had peak fluxes <0.3 Jy. Furthermore, just 3 of these subsequent detections were at velocities outside the range covered here and all 3 had peak intensities <1 Jy. So while variability is clearly an important factor in the water maser detection statistics, considerably better results are achieved by deploying an order of magnitude better sensitivity. Even so, the frequency of detections near Effelsberg's threshold suggests that still better results would obtain from a further increment in sensitivity.

Stellar velocities were known for 51 (21%) of the sample, though only 5 (10%) of these were detected here. However 20 of them have since been detected, which results in a 49% detection rate for them overall. This rate is influenced by the fact that when a circumstellar maser has already been found, more effort is devoted to the search for a water maser. But it is also noteworthy that 11 of these detections came from the first Effelsberg search, and 9 from the second, while the remaining nondetections have been examined 3-4 times without success. This suggests that it is both advantageous to have a reliable velocity, and that more detections are likely if these objects are observed again with increased sensitivity.

These data have been incorporated in past summaries of the masing status of IRAS sources (e.g. Lewis 1996), where the detection frequency as a function of (25-12) μm and/or LRS type is investigated. They have also been actively used in seeking OH masers from objects with known water masers (Lewis *et al.* 1995).

I am glad to acknowledge the help I received from Aubrey Haschick and the band of Haystack telescope operators in conducting these observations. This research made use of the SIMBAD database maintained at CDS, Strasburg, France. It was supported by NAIC, which is operated by Cornell University under a cooperative agreement with the National Science Foundation.

**Table 1: 22 GHz detections**

| **Source** IRAS | other | $S_{pk}$ (Jy) | $V_{pk}$ (km $s^{-1}$) | $\int S_\nu \, d\nu$ (Jy km $s^{-1}$) | σ (Jy) | **note** |
|---|---|---|---|---|---|---|
| 00193-4033 | BE Phe | 0.95 | -94.4 | 2.64 | 0.18 | * |
| 02145+7831 | AG Cep | 1.86 | +0.2 | 2.39 | 0.32 | * |
| 02302+4525 | UX And | 0.77 | -4.8 | 1.83 | 0.14 | * |
| 02420+1206 | RU Ari | 5.88 | +19.9 | 9.17 | 0.31 | 1,2,3 |
| 03318-1619 | RT Eri | 3.04 | +24.6 | 15.56 | 0.57 | 4 |
| 03507+3623 | IRC+40072 | 4.65 | -98.0 | 7.95 | 0.59 | 4 |
| 04157-1837 | RS Eri | 8.90 | +48.7 | 22.31 | 0.51 | * |
| 04260+2437 | IRC+20082 | 6.42 | +4.3 | 4.78 | 0.34 | 3, 4 |
| 05220-0611 | EX Ori | 4.11 | -34.7 | 3.21 | 0.30 | * |
| 06423+0905 | FX Mon | 1.51 | +34.9 | 1.40 | 0.32 | 3,4,5 |
| 13114-0232 | SW Vir | 0.93 | -8.2 | 2.62 | 0.20 | 6 |
| 13269-2301 | R Hya | 4.41 | -10.4 | 10.31 | 0.63 | ** |
| 14280-2952 | Y Cen | 2.05 | +2.0 | 9.16 | 0.41 | * |
| 14524-2148 | EG Lib | (1.51) | (+7.0) | (4.61) | 0.35 | 4 |
| 15410-0133 | BG Ser | 4.67 | -2.0 | 8.95 | 0.33 | 3 |
| 18213+0335 | IRC 00349 | 15.26 | -37.2 | 35.86 | 0.47 | 4 |
| 19146+0959 | IRC+10414 | 1.07 | -0.2 | 1.54 | 0.20 | 4 |
| 19564-0801 | RS Aql | 11.33 | +15.6 | 9.32 | 0.44 | * |
| 20248-2825 | T Mic | 1.34 | +26.9 | 7.22 | 0.39 | 3 |
| 21086+5238 | IRC+50362 | 1.89 | -10.8 | 2.07 | 0.26 | 3 |

** first detected at 22 GHz by Crocker & Hagen (1983)
1 water detection: Takaba *et al.* (1994)
2 water detection: Benson *et al.* (1996)
3 water detection: unpublished Effelsberg detection
4 OH mainline detection: Lewis *et al.* (1995)
5 1612 MHz detection: Lewis (1994)
6 SiO detection: Patel *et al.* (1992)
* no other corroboration

## Figure Legends

Figure 1: Hanning smoothed spectra of the new detections.

Figure 2: IR color distribution of the sample. Symbols: the new Haystack detections are shown as o, nondetections that have been subsequently detected at 22 GHz are shown as diamonds, while the other nondetections are shown as x. The background distribution (.) are the IRAS sources with known OH masers.

Figure 3: The S(25) v (25-12) μm distribution of the sample; the symbols are as in Fig. 2.

Table 2 Undetected IRAS sources (this 22 GHz search)

| IRAS | H2O | vel. range | IRAS | H2O | vel. range |
|---|---|---|---|---|---|
| 00186+5940 | N | (-50_+130) | 05367+3736 | N | (-95_+5) |
| 00205+5530 | N | (-63_+37) | 05368+2841 | D | (-130_+130) |
| 00222+6952 | N | (-130_+50) | 05534+4530 | N | (-130_+130) |
| 00445+3224 | N | (-50_+50) | 05559+3825 | N | (-44_+56) |
| 00484+6238 | N | (-50_+50) | 06027-1628* | N | (-130_+130) |
| 00498+4708 | N | (-50_+160) | 06088+2152 | N | (-50_+130) |
| 00541+4825 | N | (-50_+50) | 06170+3523 | N | (-50_+50) |
| 00546+5808 | N | (-130_+130) | 06439+3019 | D | (-22_+78) |
| 01080+5327 | N | (-130_+130) | 07021-0852* | D | (-50_+50) |
| 01133+2530 | N | (-50_+50) | 07042+0857 | N | (-50_+130) |
| 01149+0840 | N | (-130_+130) | 07042+2822 | N | (-130_-30) |
| 01234+5454 | N | (-130_+130) | 07152-3444* | N | (-57_+43) |
| 01246-3248 | N | (-130_+130) | 07179+2505 | N | (-130_+130) |
| 01251+1626 | N | (-50_+50) | 07245+4605 | N | (-130_+130) |
| 01280+0237 | N | (-50_+50) | 07299+0825 | N | (-1_+99) |
| 01438+1850 | N | (-50_+50) | 07422+3054 | D | (-44_+56) |
| 01472+5329 | N | (-130_+130) | 07434-3750* | N | (-50_+130) |
| 01556+4511* | D | (-130_+130) | 07487-0229 | N | (-50_+50) |
| 01562+5434 | N | (-50_+50) | 07585-1242 | D | (-68_+32) |
| 02000+0726 | D | (-50_+50) | 08063+6522 | N | (-130_+130) |
| 02028+4029 | N | (-130_+130) | 08117+2453 | N | (-50_+50) |
| 02143+4404 | N | (-130_+130) | 08196+1509 | N | (-50_+50) |
| 02251+5102 | D | (-50_+50) | 08375-1707 | D | (-90_+50) |
| 02272+3758 | D | (-50_+50) | 08437+0149 | N | (-130_+130) |
| 02380+3059 | N | (-130_+130) | 08555+1102 | N | (-50_+130) |
| 02384+3418 | N | (-90_+36) | 09019+6458 | N | (-50_+130) |
| 02455+1718 | Da | (-55_+45) | 09057+1325 | N | (-130_+130) |
| 02455-1240 | N | (-50_+50) | 09069+2527 | D | (-50_+50) |
| 02497-0828 | N | (-50_+50) | 09076+3110* | N | (-140_-40) |
| 02532+5426 | N | (-130_+130) | 10112+5635 | N | (-130_+130) |
| 02566+2938 | D | (-50_+50) | 10131+3049* | N | (-85_+15) |
| 03082+1436 | N | (-106_-06) | 10305+7001 | N | (-130_+130) |
| 03118+4623 | N | (-130_+130) | 10350-1307 | N | (-130_+130) |
| 03388-1054 | N | (-50_+50) | 10353-1145 | N | (-130_+130) |
| 03415+8010 | N | (-130_+130) | 10425-0633 | N | (-130_+130) |
| 03461+6727 | N | (-130_+130) | 10457+3633 | N | (-130_+130) |
| 03463-0710 | N | (-50_+50) | 10491-2059* | N | (-130_+130) |
| 04020-1551* | N | (-130_+130) | 11125+7524 | N | (-130_+130) |
| 04137+3114 | D | (-130_+130) | 11213-1938 | N | (-130_+130) |
| 04166+4056 | Db | (-50_+50) | 11251+4527 | N | (-130_+130) |
| 04250+1555 | N | (-130_+130) | 11461-3542* | N | (-130_+130) |
| 04265+5718 | N | (-64_+36) | 11538+5808 | N | (-130_+130) |
| 04307+6210 | N | (-64_+36) | 12020+0254 | N | (-130_+130) |
| 04328+2824 | N | (-64_+36) | 12170-1858 | N | (-130_+130) |
| 04344+3231 | N | (-130_+130) | 12277+0441 | N | (-130_+130) |
| 04560-0608 | N | (-130_+130) | 12344+2720 | N | (-130_+130) |
| 04566+5606 | D | (-64_+36) | 12345-1715 | N | (-130_+130) |
| 05090-1154 | N | (-130_+130) | 12380+5607 | Db | (-130_+130) |
| 05132+5331 | N | (-58_+42) | 13099+5638 | N | (-130_+130) |
| 05146+2521 | D | (-64_+36) | 13562-1342 | N | (-50_+50) |
| 05174-3345 | N | (-75_+25) | 13573+2801 | D | (-130_+130) |
| 05176+3502 | N | (-64_+36) | 13574+3726 | N | (-130_+130) |
| 05208-0436 | N | (-130_+130) | 14142-1612 | D | (-50_+50) |
| 05236+3200 | N | (-64_+36) | 14152-1428 | N | (-50_+130) |
| 05265-0443 | N | (-64_+36) | 14162+6701 | N | (-58_+42) |

Table 2 (continued)

| IRAS | H2O | vel. range | IRAS | H2O | vel. range |
|---|---|---|---|---|---|
| 14277+3904 | N | (-130_+130) | 19369+2823 | Dc | (-130_+130) |
| 14371+3245* | N | (-130_+130) | 19388+2855 | N | (-130_+130) |
| 14390+3147 | N | (-130_+50) | 19401+4205 | N | (-130_+130) |
| 15191-2352 | N | (-50_+130) | 19412+0337* | D | (-85_+15) |
| 15214-2244 | D | (-53_+47) | 19461+0334 | N | (-110_+50) |
| 15314+7847 | D | (-89_+12) | 19474-0744 | D | (-17_+83) |
| 15341+1515 | N | (-130_+130) | 19503+2219 | N | (-50_+90) |
| 15448+3828 | D | (-130_+130) | 19547+1848 | D | (-47_+53) |
| 15465+2818 | N | (-130_+130) | 19563+2512 | N | (-50_+130) |
| 15492+4837 | N | (-130_+130) | 20038-2722 | D | (-50_+90) |
| 15566+3609 | N | (-130_+130) | 20135+3055 | N | (-43_+57) |
| 16011+4722 | Db | (-130_+130) | 20165+3413 | N | (-130_+170) |
| 16052+4850 | N | (-110_+50) | 20268+1606 | N | (-50_+130) |
| 16060-0124 | N | (-130_+130) | 20270+3948 | N | (-50_+50) |
| 16081+2511 | D | (-60_+40) | 20350+3741 | N | (-50_+50) |
| 16308-1601 | D | (-83_+17) | 20392+1141 | N | (-50_+50) |
| 16342+6034 | N | (-130_+50) | 20431+1754 | N | (-50_+50) |
| 16473+5753 | N | (-130_+130) | 20438-0415 | N | (-50_+50) |
| 16496+1501 | N | (-130_+130) | 20440-0105 | N | (-50_+170) |
| 16534-3030 | N | (-81_+19) | 20479+0554 | N | (-50_+50) |
| 16551-0927 | N | (-130_+130) | 20502+4709 | N | (-50_+50) |
| 16574-1032 | N | (-130_+50) | 21000+8251 | N | (-50_+50) |
| 17048-1601 | D | (-86_+14) | 21008+5930 | N | (-50_+50) |
| 17072+1844 | D | (-130_+130) | 21027+3704 | N | (-50_+50) |
| 17086+2739 | N | (-130_+130) | 21028+2711 | N | (-50_+50) |
| 17115+1803 | N | (-50_+130) | 21044-1637* | N | (-50_+90) |
| 17123-2122 | N | (-50_+50) | 21088+6817 | D | (-50_+50) |
| 17139+0446 | D | (-130_+130) | 21341+4508 | N | (-50_+90) |
| 17265-0725 | N | (-50_+50) | 21358+7823 | N | (-50_+50) |
| 17328-0118 | N | (-130_+130) | 21389+5405 | N | (-50_+90) |
| 17329+5359 | N | (-50_+130) | 21419+5832 | Da | (-50_+90) |
| 17342+3127 | N | (-130_+130) | 21439-0226* | D | (-50_+90) |
| 17488-2800 | N | (-130_+130) | 21563+5630 | N | (-90_+50) |
| 18040-0941* | N | (-50_+50) | 22000+5643 | N | (-130_+130) |
| 18157+1757 | N | (-130_+130) | 22036+3315 | N | (-130_+130) |
| 18181+2550 | N | (-130_+130) | 22039+6215 | N | (-50_+130) |
| 18448-0545 | N | (-130_+130) | 22233+3013 | D | (-75_+25) |
| 18479+0432 | N | (-130_+130) | 22306+5510 | N | (-10_+90) |
| 18501-2132 | D | (-50_+50) | 22395+4217 | N | (-50_+130) |
| 18520-1635 | N | (-50_+50) | 22406+2753 | N | (-50_+130) |
| 18537-1035 | N | (-130_+130) | 22456+5453 | N | (-63_+90) |
| 19007-2247 | N | (-130_+130) | 22476+4047 | N | (-63_+90) |
| 19008+0726 | N | (-130_+130) | 22553+1744 | Db | (-90_+90) |
| 19071+2934 | D | (-63_+37) | 22572+4234 | N | (-90_+90) |
| 19099+6711 | Dc | (-130_+130) | 23025+6429 | N | (-50_+130) |
| 19123+0409 | N | (-50_+50) | 23095+5925 | N | (-79_+21) |
| 19126-0708* | N | (-50_+50) | 23180+0838 | N | (-50_+50) |
| 19129+2803 | N | (-130_+130) | 23201-1105 | N | (-50_+130) |
| 19147+5004 | N | (-130_+130) | 23320+4316 | N | (-63_+90) |
| 19175-0807 | N | (-50_+50) | 23412-1533 | N | (-90_+23) |
| 19194+1734 | N | (-50_+50) | 23420+5618 | N | (-90_+20) |
| 19232+5008 | N | (-50_+50) | 23554+5612 | N | (-50_+130) |
| 19240+1634 | N | (-130_+150) | 23575+2536 | N | (-50_+130) |
| 19287+4602 | N | (-130_+130) | | | |
| 19296+4331 | N | (-130_+170) | | | |

* first 22 GHz search
a Takaba et al. (1994)
b Szymczak & Engels (1995)
c Benson et al. (1996)

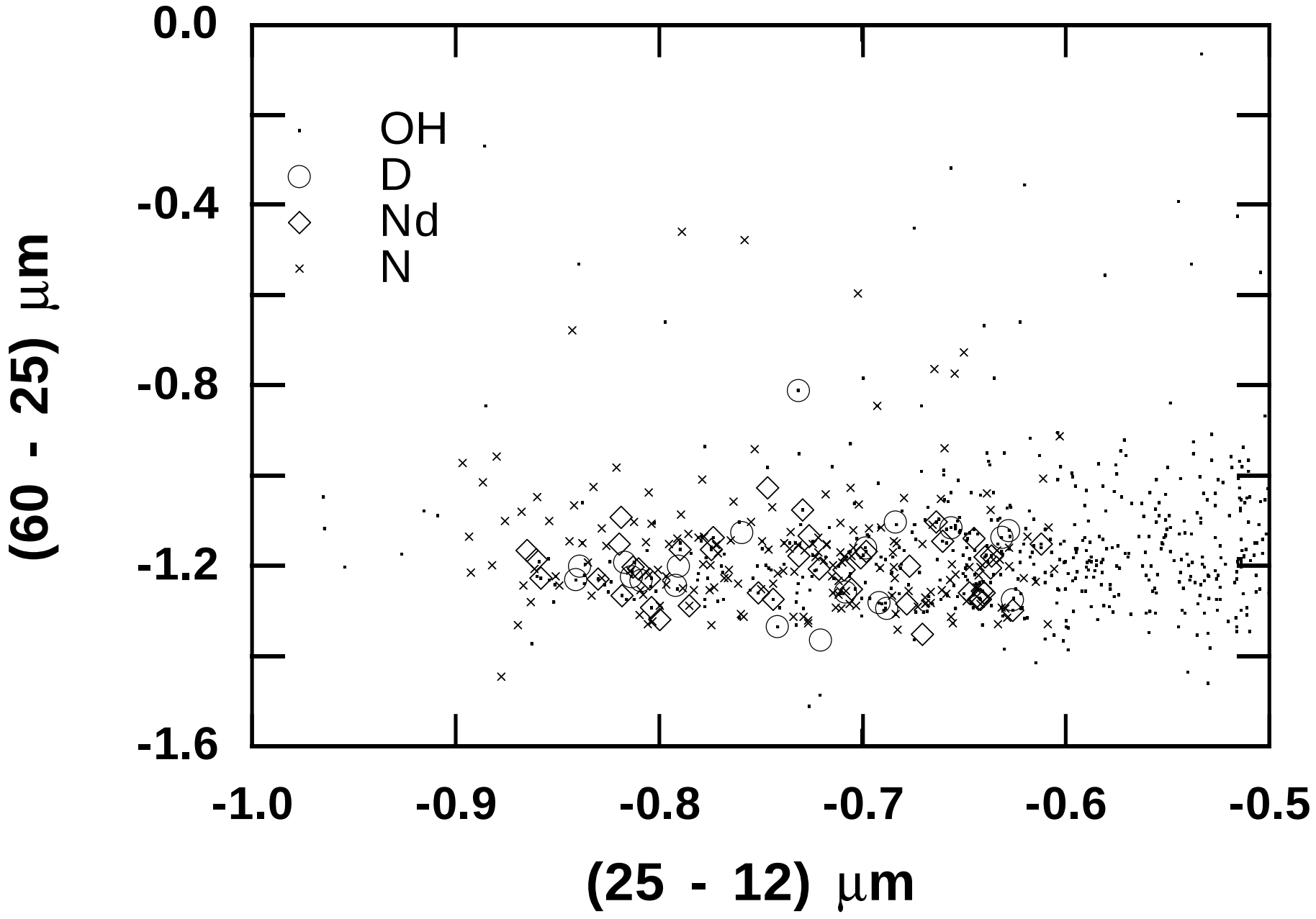

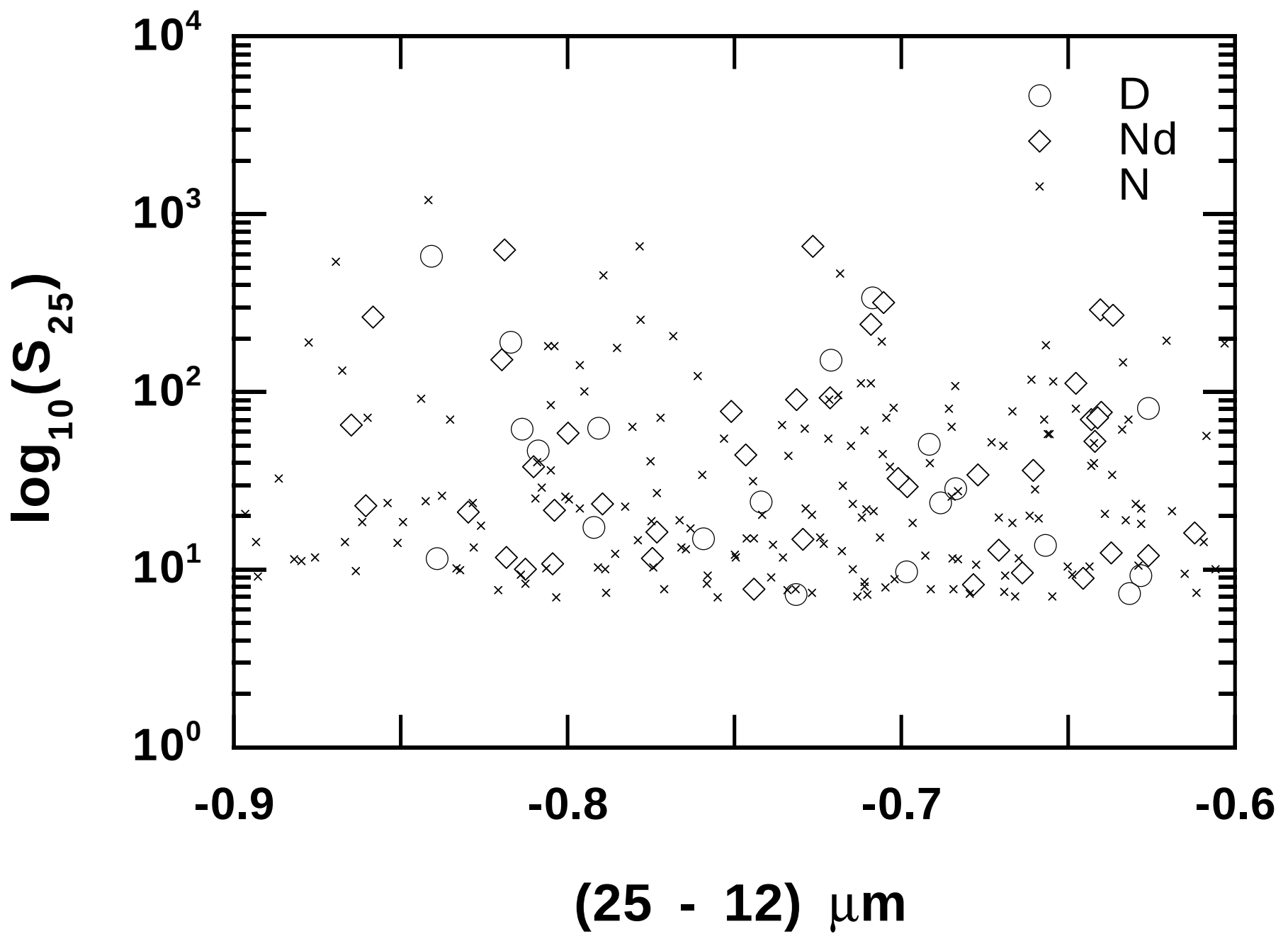